\documentstyle[preprint,version2,aps]{revtex}

\begin{document}
\begin{title}
Superconductivity in the Anderson Lattice: Effects
of Crystal Fields on Quasiparticle Interactions
\end{title}
\author{B. R. Trees and D. L. Cox}
\begin{instit}
Department of Physics, Ohio State University, Columbus, OH 43210
\end{instit}

\begin{abstract}
Using an anisotropic hybridization matrix element, V($\vec k$),
reflecting cubic symmetry at the ``rare-earth'' sites of the
infinite U Anderson lattice, we calculate the bare quasiparticle
scattering amplitude to order 1/N using slave Bosons and find significant
differences with previous results under spherical symmetry.
Due to exchange of density
fluctuations, we find an instability.
toward E$_g$ and T$_{1g}$ (even-parity) pairing.
\end{abstract}
\pagebreak
\narrowtext

Since the advent of heavy Fermion superconductivity as discovered in
CeCu$_2$Si$_2$[1], physicists have puzzled over the cause and
nature of the pairing in these materials.  Unique for several
reasons, heavy Fermion compounds are comprised of intermetallics and
rare-earth or actinide atoms (such as U or Ce) characterized by:  partially
filled 4f or 5f shells; a large local Coulomb repulsion; and
a rich multiplet structure in the presence of crystal electric fields.
Hybridization between
conduction electrons and these cerium or uranium electrons gives rise
to extremely heavy quasiparticles which participate in the pairing [1].
These properties, coupled with the fact that the quasiparticle bandwidth
(set by the lattice Kondo temperature T$_o$) is much less than the
Debye temperature, make heavy Fermions strong
candidates for non-phononic, anisotropic superconductors [2],[3].
Our work focuses on calculating the two-particle scattering amplitude for
CeCu$_2$Si$_2$, and we see that the inclusion of
crystal electric fields of cubic symmetry
at the Ce sites strongly modifies the anisotropy
and strength of the quasiparticle interactions, leading to a superconducting
instability of either E$_g$ or T$_{1g}$ symmetry.

Most agree that a lattice of local electrons on ``rare-earth'' sites hybridized
with
conduction electrons captures the essential physics of heavy
Fermions in the normal state. Anderson suggested [4]
that the inverse degeneracy of the ``rare-earth'' multiplet would
be a useful expansion parameter to quantify the strength of
quasiparticle interactions in the lattice.  In the limit of infinite Coulomb
repulsion, Coleman [5] made a low temperature (T$<$T$_o$),
strong-coupling calculation
tractable by introducing slave bosons, b($\vec R$$_i$), (as first used
by Barnes [6]) to represent a hole or empty site on the lattice.
A physical electron on a rare-earth site is then created by the product of
a fermionic operator
and a slave boson:  f$^\dagger$$_{\Gamma \alpha}$($\vec R$$_i$)b($\vec R$$_i$),
where
$\Gamma$ and $\alpha$ are crystalline quantum numbers at site i.  There
is also a Lagrange multiplier
to constrain the
number of Fermions and Bosons
at each site to unity.

Lavagna, Millis, and P. Lee [7] looked for a superconducting
instability in the Anderson lattice using this formalism.
Within the so-called SU(N) version of the model,
where one assumes N-fold
degenerate conduction states, and with a spherically symmetric hybridization
matrix element V, they found a d-wave pairing instability at order 1/N.
Zhang and T. K. Lee [8] pointed out the need to include
spin-orbit coupling and an anisotropic hybridization in order to model
heavy Fermion systems realistically.  They showed that even-parity
pairing channels had large local, or ``hard-core'', repulsions which
prevented pairing.

Our work reveals that the local repulsive interactions are
substantially changed due to the cubic crystal fields, and this change is in
large
part responsible for the pairing instability.
The result thus can be seen as a natural evolution of previous calculations
[7,8],
wherein the crystal field split multiplet structure is, in some sense,
the last major piece of local physics to be added to the model.

CeCu$_2$Si$_2$ has a superconducting transition temperature of about
0.6 K and a Kondo temperature (effective quasiparticle bandwidth) of
approximately 10 K  (T$_o$$\approx$10K) [1].
Inelastic neutron scattering measurements by Horn
{\it et al.}[9] clearly show the crystal field splitting of the low-lying
Ce 4f$^1$ spin-orbit multiplet (J=5/2). Although CeCu$_2$Si$_2$ has a
tetragonal unit cell, fits to specific heat [10],static susceptibility
[11], and de-Haas van-Alphan data for the effective mass[12]
 require that the Ce ions actually occupy sites of pseudo-cubic
symmetry. Thus, our work focuses on plane wave
conduction states hybridizing with localized crystal fields states of cubic
symmetry.

In the presence of crystal electric fields of cubic symmetry, a J=5/2
multiplet is split into a doublet of $\Gamma$$_7$ symmetry and a quartet of
$\Gamma$$_8$ symmetry [13].  From the inelastic neutron scattering
data of reference 9, we know that the doublet has the lower energy in
CeCu$_2$Si$_2$ with the excited quartet about 36 meV higher
(i. e. $\Delta$$_{CEF}$=36 meV $\approx$ 360 K).  These states can be expressed
as a linear combination of the J=5/2 states as shown in reference 13, and
our hybridization matrix elements can be written as
$$V_{\Gamma \alpha \sigma}(\vec k)=\sum_{m=-\frac{5}{2}}^{\frac{5}{2}}c_{\Gamma
\alpha m}V_{m \sigma}(\vec k), \eqno (1a)$$
where $\alpha$ labels the degenerate states for a given representation
$\Gamma$ and
the expansion coefficients c$_{\Gamma \alpha m}$ are given in
reference 13. The mixing matrix element
V$_{m \sigma}$ has the Coqblin-Schrieffer form
$$V_{m \sigma}(\vec k)=(-i)^3\sigma V_k\sqrt{\frac{7-2m\sigma}{14}}Y^*_{3,m-
\frac{\sigma}{2}}(\hat k),\eqno (1b)$$
where $\sigma$=$\pm$1 is the spin index.

We build upon the formalism of Read and Newns[14] and calculate the partition
function as a functional integral over Grassmann and complex fields.
The appropriate action can ultimately be constructed from the Hamiltonian:
$$H=\sum_{\vec k \sigma} \xi_kc^\dagger_{\vec k \sigma}c_{\vec k \sigma}
+\sum_{\vec k \Gamma \alpha}E_\Gamma f^\dagger_{\vec k \Gamma \alpha}
f_{\vec k \Gamma \alpha}$$
$$+\sum_{\vec k \vec q \Gamma \alpha \sigma}\biggl[\tilde V_{\Gamma \alpha
\sigma}
(\vec k)c^\dagger_{\vec k \sigma}f_{\vec k + \vec q \Gamma \alpha}\tilde s_{
\vec q \Gamma}+ c. c.\biggr]$$
$$\sum_{\vec k \vec q}i\lambda_{\vec q}\biggl[\sum_{\Gamma \alpha}f^\dagger_{
\vec k + \vec q \Gamma \alpha}f_{\vec k \Gamma \alpha}+\frac{1}{2}\sum_{\Gamma}
\biggl(N_{\Gamma}\tilde s_{\vec k + \vec q \Gamma}\tilde s_{\vec k \Gamma}-
N_{\Gamma}q_{-\vec q \Gamma}\biggr)\biggr],\eqno (2)$$
where c$_{\vec k \sigma}$ and f$_{\vec k \Gamma \alpha}$ destroy, respectively,
a conduction electron of crystal momentum $\vec k$ and spin $\sigma$ and an
electron in a multiplet state with crystal field quantum numbers $\Gamma$ and
$\alpha$. $\xi$$_{\vec k}$ is the quadratic conduction electron
dispersion.  E$_\Gamma$ represents the bare crystal field energies,
N$_{\Gamma}$
is the degeneracy of the $\Gamma$ multiplet, and q$_{\Gamma}$=1/N$_{\Gamma}$.
The last term is the constraint of unit occupancy at each Ce site, and
all energies are measured relative to the chemical potential for the free
conduction electrons.

Following Read and Newns, $\tilde s$$_{\vec q \Gamma}$ is the
amplitude of a complex field which, at mean-field level, renormalizes the
hybridization strength, $\tilde V$$_{\Gamma\alpha}$.  One can relate
{\em s}$_\Gamma$ to
Coleman's slave Boson via {\em b}$_\Gamma$={\em s}$_\Gamma${\em
e}$^{(i\theta)}$,
where the phase, $\theta$,
can be absorbed into the Fermion via a gauge transformation.
Notice the rescaled
quantities $\tilde V$$_{\Gamma \alpha \sigma}$=$\sqrt{
N_\Gamma}$V$_{\Gamma \alpha \sigma}$ and $\tilde s$$_{\vec k \Gamma}$=s$_{\vec
k
\Gamma}$/$\sqrt{N_\Gamma}$, where $\tilde V$$_{\Gamma\alpha}$ and
$\tilde s$$_\Gamma$
are assumed of order
one. As elucidated in reference 14, by choosing this so-called radial
gauge, the theory is not plagued by unphysical infrared divergences.

At the mean-field level, we assume the {\it s$_o$} and {\it
$\lambda$$_o$} fields are
uniform in space and time, integrate out the Fermionic degrees of freedom and
require the resulting free energy be stationary with
respect to {\it s$_o$} and {\it $\lambda$$_o$}. For simplicity, we have assumed
that
$\tilde s$$_{o\Gamma}$={\it s$_o$}/$\sqrt{N_\Gamma}$, where {\it s$_o$} is the
same
for both crystal field multiplets. We solve the resulting equations,
along with an equation for the quasiparticle chemical potential $\mu$,
self-consistently.  See Table 1 for a set of mean-field parameters that
should be applicable to CeCu$_2$Si$_2$.

The quasiparticle states can of course be written as a combination of plane
waves and crystal field states:
$$|\psi_{n \sigma}(\vec k)>=A_n(\vec k)\biggl[|\vec k \sigma>-
\sum_{\Gamma \alpha}\frac{\tilde s_{o\Gamma}\tilde V_{\Gamma \alpha
\sigma}^*(\vec k)}{\epsilon_\Gamma-E_n(\vec k)}|\Gamma \alpha>
\biggr], \eqno(6a)$$
where $\epsilon$$_{\Gamma}$=E$_{\Gamma}$+i$\lambda$$_o$ are the energies
of the shifted crystal field levels and E$_n$($\vec k$) is a quasiparticle
energy for the nth band.
The normalization factor
$$A_n(\vec k)^2=\biggl[1+\sum_{\Gamma \alpha}\frac{\tilde s_{o\Gamma}^2
|\tilde V_{\Gamma \alpha \sigma}(\vec k)|^2}{\bigl(\epsilon_\Gamma-E_n(\vec k)
\bigr)^2}\biggr]^{-1}  \eqno(6b)$$
is highly anisotropic.

It is important to note
that the work of Zhang and T. K. Lee, even in the presence of
anisotropic hybridization, is based on overall spherical symmetry at the
cerium ions.  Thus, their normalization function is isotropic.  In
our case, the constraints of cubic symmetry lead to the strong anisotropies.
Specifically, the hybridization with localized states of $\Gamma$$_7$
symmetry vanishes along special directions in k space, resulting in
a sharp peak in A$_1$($\vec k$) which would actually diverge if the
hybridization with $\Gamma$$_8$ states were not nonzero.  This
enhanced normalization affects quasiparticle interactions significantly,
as will be discussed below.

Due to the two Bosonic fields, {\em s}$_o$ and $\lambda$$_o$, the
dressed Bosonic propagator
is conveniently calculated from a matrix Dyson's equation
D$_{ab,\Gamma}$$^{-1}$($\vec
q$)=D$^{(o)}$$_{ab,\Gamma}$$^{-1}$-$\Pi$$_{ab,\Gamma}$
($\vec q$).
The bare propagator, D$^{(o)}$$_{ab,\Gamma}$, is of order 1/N$_\Gamma$,
and taking only
closed Fermion loops for the self-energy insures that the dressed propagator is
also of order 1/N$_\Gamma$.
Allowing the Bose fields
to fluctuate away from the mean-field values and expanding the
free energy up to terms quadratic in these fluctuations, we can
calculate the self-energy $\Pi$$_{ab,\Gamma}$($\vec q$), which in the static
limit has the following general susceptibility-like structure:
$$\Pi_{ab,\Gamma}(\vec q)=P\sum_{\vec k \vec k^\prime}\frac{M_{ab,\Gamma}(\vec
k,
\vec k^\prime)}{E_1(\vec k^\prime)-E_1(\vec k)}\delta_{\vec k^\prime,
\vec k \pm \vec q + \vec Q}.  \eqno (7)$$
The effective ``matrix element'', M$_{ab,\Gamma}$($\vec k$,$\vec k$$^\prime$),
is a complicated anisotropic function, and $\vec Q$ is a reciprocal lattice
vector.

The principal value nature of this Brillouin zone sum makes it
challenging.  We proceed by borrowing
a numerical technique from electronic structure theory
for calculating susceptibilities:  the
analytic tetrahedron method [15].  In this procedure we break up the
Brillouin zone into a large number of tetrahedra, over which the integral of
1/(E$_1$($\vec k$$^\prime$)-E$_1$($\vec k$)) can be performed essentially
analytically.  We assume the matrix element is constant inside a given
tetrahedron, and
since M$_{ab,\Gamma}$($\vec k$$^\prime$,
$\vec k$) is a complicated function of $\vec k$ and $\vec k$$^\prime$,
this produces fluctuations in the result as a function of mesh size.
This is the most time consuming part of the calculation.
Our final results for the effective quasiparticle interactions
will be presented as an average over many different self-energy
meshes and will include a standard deviation reflecting the fluctuations
involved.
We wish to emphasize,
however, that this ``matrix element effect'' is a well known problem in
electronic structure calculations and the current approach
represents the state of the art [16].

The particle-particle scattering amplitude in the even-parity
channel is shown in figure 1.  We believe even-parity pairing in CeCu$_2$Si$_2$
is suggested by the following experimental evidence:
  the necessity of strong Pauli limiting to describe the low temperature
behavior of the upper critical field H$_{c2}$(T) (odd-parity would
probably produce only a small Pauli limiting)
 [17]; appearance of a
large DC-Josephson current through a CeCu$_2$Si$_2$/Al weak-link. (At
the simplest level, Al is a singlet BCS superconductor and it is difficult
to get a Josephson current between opposite parity superconductors.)[18];
 and the strong reduction of the $^{29}$Si and $^{63}$Cu Knight shifts
below T$_c$ [19].

Because of computational complexity, we confine our calculation to the
Fermi surface.  This weak-coupling calculation
means, of course, that we miss any evidence of a pairing state with
unconventional frequency dependence (a topic which has grown in
interest recently).  We also miss any strong-coupling renormalization
of the effective interaction.  This should not jeopardize, however,
our results for instabilities in the E$_g$ and T$_{1g}$ pairing
channels, since they depend largely on the underlying symmetries
of the model and would be material dependent.

For simplicity, we have assumed a spherical Fermi surface.  For very small
fillings of the lowest quasiparticle band (n$_{total}$=n$_c$+n$_f$$\ll$1)
the Fermi surface is indeed truly spherical.  Only for a nearly filled
band (n$_{total}$$\approx$ 2) should the surface be significantly warped,
reflecting the $\Gamma$$_7$ symmetry of the problem.  We take a filling
(n$_{total}$=1.5) that gives a Fermi surface only mildly distorted from a
sphere and yet still leads to a large quasiparticle density of states.

We have the following form for the particle-particle scattering amplitude (see
reference 8):
$$\Gamma_o\bigl(\vec k,\vec k^\prime\bigr)=\frac{1}{4}\sum_{\Gamma \alpha
\Gamma^\prime \alpha^\prime \sigma \sigma^\prime}\frac{\tilde s_{o \Gamma}
\tilde s_{o \Gamma^\prime}}{T_{o \Gamma}T_{o \Gamma^\prime}}
\tilde V_{\Gamma \alpha \sigma}^*(\vec k)\tilde V_{\Gamma^\prime
\alpha^\prime \sigma}(\vec k)$$
$$\times\tilde V_{\Gamma^\prime \alpha^\prime
\sigma^\prime}^*(\vec k^\prime)\tilde V_{\Gamma \alpha \sigma^\prime}
(\vec k^\prime)
\times A_1(\vec k_F)^2A_1(\vec k_F^\prime)^2\times $$
$$\biggl[D_s^{(o)}-iD_{s\lambda}^{(o)}
\biggl(\frac{\tilde s_{o\Gamma}}{T_{o\Gamma}}+\frac{\tilde s_{o\Gamma^\prime}}
{T_{o\Gamma^\prime}}\biggr)\biggr]\biggl[\frac{1}{1+F(\vec k^\prime-\vec k)}
+\frac{1}{1+F(\vec k^\prime+\vec k)}\biggr],  \eqno (7)$$
where T$_{o\Gamma}$ is an effective Kondo scale for the $\Gamma$ crystal
field multiplet (T$_{o\Gamma}$=$\epsilon$$_{\Gamma}$-$\mu$),
D$_s$$^{(o)}$=-$\tilde s$$_{o\Gamma_7}$$^2$/2T$_{o\Gamma_7}$,
D$_{s\lambda}$$^{(o)}$=i$\tilde s$$_{o\Gamma_7}$, and D$_\lambda$$^{(o)}$=0.
Recall that A$_1$($\vec k$) is the anisotropic quasiparticle
normalization.  The function F($\vec k$$^\prime$ $\pm$ $\vec k$) has
a complicated dependence on the Bosonic self-energy.

For our weak coupling calculation it remains to average the scattering
amplitude over the Fermi surface
$$<\Gamma_o>_a=\int \frac{d\hat k}{4\pi}\int \frac{d\hat k^\prime}{4\pi}
\phi_a^*(\hat k^\prime)\Gamma_o(\hat k^\prime,\hat k)\phi_a(\hat k),
\eqno (8)$$
where the $\phi$$_a$($\hat k$) are the so-called cubic harmonics, i. e.
functions that transform under the representations of the group O$_h$.
The subscript a labels these representations (A$_{1g}$,A$_{2g}$,E$_g$,
T$_{1g}$,T$_{2g}$).  A superconducting instability in the ``a-th'' pairing
channel is signaled by a negative value of the corresponding average.

We now present our main results.  First, it is instructive to consider the
limit F($\vec k$$^\prime$$\pm$$\vec k$)$\rightarrow$ 0 (see Table 1).
This leaves us with
only the local interactions between the quasiparticles, which have a large
repulsive strength in the A$_{1g}$ (s-wave like) channel and are enhanced by
the
strong anisotropy of the normalization function.
Furthermore, a combination of the normalization
and the hybridization matrix element causes the local
interactions to be {\it attractive} in the E$_g$(d$_{x^2-y^2}$) channel
and repulsive but small in the T$_{1g}$ and T$_{2g}$ channels.

Our results contrast with those of Zhang and Lee[8].  They find repulsive
local interactions in the s,d, and g wave pairing channels that are all
approximately the same size. We emphasize that
our new results stem from the lowering
of the symmetry from spherical to cubic.

The last step is to include the full bosonic self-energy through the
function F($\vec k$$^\prime$$\pm$$\vec k$),
which is strongly anisotropic due to its dependence on the effective matrix
element, M$_{ab,\Gamma}$.  This anisotropy gives rise to an
averaged scattering amplitude (equation 8) that varies with the
self-energy mesh size.
We have thus averaged the results of equation 8 over eleven different
self-energy meshes, and the results for the effective quasiparticle
interactions
are:
$$\lambda_{A_{1g}}=-<\Gamma_o>_{A_{1g}}/T_{o\Gamma_7}=-1.70 \pm 0.04$$
$$\lambda_{E_g}=-<\Gamma_o>_{E_g}/T_{o\Gamma_7}=0.133 \pm 0.10$$
$$\lambda_{T_{1g}}=-<\Gamma_o>_{T_{1g}}/T_{o\Gamma_7}=0.231 \pm 0.18$$
$$\lambda_{T_{2g}}=-<\Gamma_o>_{T_{2g}}/T_{o\Gamma_7}=-0.616 \pm 0.45.$$

\noindent
The fluctuations may be large because of the matrix elements in the
Bosonic self-energy,
but a superconducting instability in the
E$_g$ and T$_{1g}$ channels is still clear.

We wish to remark that a strong-coupling calculation, including the
dynamics of the exchanged boson, is currently intractable with this procedure.
However, we don't expect it to modify the qualitative aspects of our stability
analysis since the quasiparticles interactions at $T_c$ are renormalized by
factors of $1/(1+\lambda)$ where $\lambda$ arises from the frequency dependence
of the quasiparticle self energy.  Since the quasiparticle self energy in
heavy fermion materials is expected to be largely momentum independent, this
shouldn't affect the ordering of the $\lambda_{\Gamma}$ couplings.
Furthermore, because of the complexity of reliably computing
the self-energy, pushing even
the static calculation to order 1/N$^2$ also appears unrealistic.  The
anisotropies due to the cubic symmetry have increased the complexity
of the problem more than one might have expected.  We do imagine, however,
studying other Ce based systems such as CeAl$_3$ at the 1/N level [20].

It is a pleasure to acknowledge
helpful conversations with M. Alouani, M. Steiner, and J. W. Wilkins.
This research has been supported by a grant from the U. S. Department of
Energy, Office of Basic Energy Sciences, Division of Materials Research.

\pagebreak

{\bf Figure:} Quasiparticle scattering amplitude in the pseudospin
singlet channel.  The
full lines represent quasiparticles at the Fermi surface; the dashed line
is the dressed boson propagator.  The large circles are bare, anisotropic
vertices that connect quasiparticles with the bosons.

\pagebreak

\begin{table}
\caption{Fermi Surface Average of the Local ``Hard-Core'' Quasiparticle
Scattering Amplitude.
``a'' labels the
irreducible representations of the group O$_h$. ``local''
means the bosonic self-energy was not included.  The results from
Zhang and Lee[8] are also presented for comparison.  It is clear that
cubic symmetry yields different quasiparticle interactions than spherical
symmetry.
The mean-field parameters used in our calculations are:
Kondo temperature (T$_{o7}$=1.30 meV); crystal
field splitting ($\Delta$$_{CEF}$=0.036 eV); hybridization
renormalization (s$_o$=0.1424094); and bare hybridization strength
(V$_o$=0.65 eV).}
\end{table}
\pagebreak

\begin{tabular}{|c|c||c|c|} \hline
\multicolumn{2}{|c||}{Cubic Crystal Fields} &
   \multicolumn{2}{|c|}{Spherical Symmetry (from [8])} \\ \hline
a&$\langle$$\Gamma$$_o$$\rangle$$_a$/T$_{o\Gamma}$$_7$&l&
$\langle$$\Gamma$$_o$$\rangle$$_l$/T$_o$ \\ \hline
A$_{1g}$&3.79&l=0(s-wave)&1/6 \\
E$g$&-0.019&l=2(d-wave)&4/21 \\
T$_{1g}$&0.130&l=4(g-wave)&1/7 \\
T$_{2g}$&0.133& & \\ \hline
\end{tabular}

\end{document}